\documentclass[11pt]{article}
\usepackage{graphicx}

\begin{document} 

\def\xslash#1{{\rlap{$#1$}/}}
\def \p {\partial}
\def \dd {\psi_{u\bar dg}}
\def \ddp {\psi_{u\bar dgg}}
\def \pq {\psi_{u\bar d\bar uu}}
\def \jpsi {J/\psi}
\def \psip {\psi^\prime}
\def \to {\rightarrow}
\def \lrto{\leftrightarrow} 
\def\bfsig{\mbox{\boldmath$\sigma$}}
\def\DT{\mbox{\boldmath$\Delta_T $}}
\def\xit{\mbox{\boldmath$\xi_\perp $}}
\def \jpsi {J/\psi}
\def\bfej{\mbox{\boldmath$\varepsilon$}}
\def \t {\tilde}
\def\epn {\varepsilon}
\def \up {\uparrow}
\def \dn {\downarrow}
\def \da {\dagger}
\def \pn3 {\phi_{u\bar d g}}

\def \p4n {\phi_{u\bar d gg}}

\def \bx {\bar x}
\def \by {\bar y}

\begin{center}
{\Large\bf Scale Dependence of Twist-3 Quark-Gluon Operators for 
 Single Spin Asymmetries}
\par\vskip20pt
J.P. Ma$^{1,2}$  and Q. Wang$^{3}$    \\
{\small {\it $^1$ Institute of Theoretical Physics, Academia Sinica,
P.O. Box 2735,
Beijing 100190, China\\
$^2$ Center for High-Energy Physics, Peking University, 100871, China  \\
$^3$ Department of Physics and Institute of Theoretical
Physics, Nanjing Normal University, Nanjing, Jiangsu 210097, P.R.China }}
\end{center}
\vskip 1cm
\begin{abstract}
We derive the scale dependence of twist-3 quark-gluon operators, or ETQS matrix elements, at one-loop. 
These operators are used to factorize transverse single spin asymmetries, which are studied intensively 
both in experiment and theory. The scale dependence of two special cases are particularly interesting.  
One is of soft-gluon-pole matrix elements, 
another is of soft-quark-pole matrix elements. From our results the evolutions in the two cases can be obtained. 
A comparison with existing results of soft-gluon-pole matrix elements is made.

\vskip 5mm \noindent
\end{abstract}
\vskip 1cm

\par 
In high energy scattering with a transversely polarized hadron transverse Single Spin Asymmetry(SSA) can appear.
Such an asymmetry has been observed in various experiments. Reviews about this research field can be 
found in \cite{Review}. The study of SSA plays an important role in exploring 
the inner-structure of hadrons. In the framework of collinear factorization, SSA can be factorized 
with matrix elements defined with twist-3 QCD operators, called ETQS matrix elements\cite{EFTE,QiuSt}. 
These matrix elements contain important information about correlations between more than two partons 
in the transversely polarized hadron and can be extracted from experimental data of SSA. 
\par 
To precisely extract the twist-3 matrix elements or predict SSA, one needs to know 
the scale dependence of the matrix elements. This dependence plays a similar role like
that of standard twist-2 parton distributions functions. The later is given by the famous DGLAP equations.  
In this letter we study the scale dependence of 
these twist-3 matrix elements. Especially, we focus on the matrix elements defined with twist-3 quark-gluon operators. 
\par  
The definitions of the interested twist-3 matrix elements can conveniently be given
with a light-cone coordinate system, in which a
vector $a^\mu$ is expressed as $a^\mu = (a^+, a^-, \vec a_\perp) =
((a^0+a^3)/\sqrt{2}, (a^0-a^3)/\sqrt{2}, a^1, a^2)$ and $a_\perp^2
=(a^1)^2+(a^2)^2$. In the system we introduce two light-cone vectors: $n^\mu =(0,1,0,0)$ and $l^\mu =(1,0,0,0)$. 
The anti-symmetric tensor $\epsilon_\perp^{\mu\nu}$ in the transverse space is defined as $\epsilon_\perp^{\mu\nu} =\epsilon^{\alpha\beta\mu\nu}l_\alpha n_\beta$  with the convention $\epsilon^{0123}=1$. There are 6 twist-3 operators
or 6 twist-3 matrix elements for a transversely polarized hadron with the momentum $P^\mu =(P^+,0,0,0)$
and the spin vector $s^\mu =(0,0, s_\perp^1,s_\perp^2)$.  
Two of them are defined with product of two quark operators and one gluon field strength operator. 
They are defined as\cite{EFTE,QiuSt}:
\begin{eqnarray}
T_F (x_1,x_2, \mu)  & =&   -\tilde s_\mu  g_s \int \frac{dy_1 dy_2}{4\pi}
   e^{ -i P^+ (y_2 (x_2-x_1)  + y_1 x_1) }
   \langle P, s \vert
           \bar\psi (y_1n )  \gamma^+  G^{+\mu}(y_2n) \psi(0) \vert P, s \rangle, 
\nonumber\\
T_{\Delta,F} (x_1,x_2)
   & =&    i s_\mu  g_s \int \frac{dy_1 dy_2}{4\pi}
   e^{ -i P^+ (y_2 (x_2-x_1)  + y_1 x_1)}
 \langle P, s \vert
           \bar\psi (y_1n )  \gamma^+ \gamma_5   G^{+\mu}(y_2n)  \psi(0) \vert P, s\rangle
\label{tw3}
\end{eqnarray}
with $\mu=1,2$ and $\tilde s^\mu = \epsilon^{\mu\nu} s_\nu$. 
The above definitions are given in the light-cone gauge $n\cdot G=0$. 
In other gauges, gauge links along the direction $n$ should be supplemented to make 
them gauge invariant. 
From general principles one can derive the following properties: 
\begin{equation} 
  T_F(x_1,x_2) = T_F(x_2,x_1), \ \ \  T_{\Delta,F} (x_1,x_2) =- T_{\Delta,F} (x_2,x_1). 
\label{TFSY} 
\end{equation}   
\par 
The remaining four twist-3 matrix elements are defined only with gluon field strength operators.
These purely gluonic matrix elements have been given in \cite{Ji3G}. 
Two of the four matrix elements are defined with the $SU(N_C)$-gauge group constant $f^{abc}$:  
\begin{eqnarray} 
T^{(f)}_{G} (x_1,x_2)  &=& -\tilde s_\mu g_s \frac{ i f^{abc} g_{\alpha\beta} }{P^+} 
      \int\frac{d y_1 d y_2}{4\pi} 
   e^{-i P^+ (y_2 (x_2-x_1) + y_1 x_1)}
\nonumber\\   
   &&   \ \ \ \ \ \ \ \  \cdot \langle P, s\vert G^{a,+\alpha}(y_1 n) G^{b, +\mu}(y_2 n) G^{c,+\beta} (0) \vert P, s \rangle, 
\nonumber\\
 T^{(f)}_{\Delta,G} (x_1,x_2)  &=& -s_\mu g_s \frac{ i f^{abc} \epsilon_{\perp\alpha\beta} }{P^+} 
     \int\frac{d y_1 d y_2}{4\pi} 
   e^{-i P^+ (y_2 (x_2-x_1) + y_1 x_1)}
\nonumber\\   
  &&  \ \ \ \ \ \ \ \  \cdot \langle P, s \vert G^{a,+\alpha}(y_1 n) G^{b, +\mu}(y_2 n) G^{c,+\beta} (0) \vert P, s \rangle.  
\end{eqnarray}
The definitions of other two $T^{(d)}_{G}$ and $T^{(d)}_{\Delta,G}$ are obtained by replacing $if^{abc}$ 
with $d^{abc}$. Again, the above definitions are given in the $n\cdot G=0$ gauge. For the matrix elements
with $f^{abc}$  one has:
\begin{eqnarray}
T^{(f)}_{G}(x_1,x_2) &=& -T^{(f)}_G(-x_2,-x_1), \ \ \ T^{(f)}_G(x_1,x_2) = T^{(f)}_G(x_2,x_1), 
\nonumber\\
T^{(f)}_{\Delta,G }(x_1,x_2) &=& T^{(f)}_{\Delta,G }(-x_2,-x_1), \ \ \ \  T^{(f)}_{\Delta,G }(x_1,x_2)=-T^{(f)}_{\Delta,G } (x_2,x_1).
\label{TSY} 
\end{eqnarray}
Similar relations for $T^{(d)}_{G}$ and $T^{(d)}_{\Delta,G}$ can also be derived. The defined six twist-3 matrix elements 
depend on the renormailzation scale $\mu$.   
\par
The support of the defined matrix elements can be analyzed with translational covariance. 
One easily obtains that the matrix elements are only nonzero for 
$\vert x_{1,2}\vert \leq 1$ and $\vert x_1-x_2\vert \leq 1$. 
There are two special cases which are particularly interesting. One is with $x_1=x_2$. The corresponding 
nonzero matrix elements are called soft-gluon-pole matrix elements, which describe the correlation 
between partons and one gluon with zero momentum fraction.  Another case is with $x_1=0$ or $x_2 =0$. 
The corresponding matrix elements are called soft-quark-pole matrix elements.
The matrix elements in two cases are of particular interesting\cite{QiuSt,SQP}.   
The scale dependence of soft-gluon-pole matrix elements has been studied in \cite{KQ,BMP,VoYu,ZYL}. But the obtained results 
are not completely in agreement. 
The renormailization of the above twist-3 operators in space-time has been studied in \cite{BB}. 
The evolution of twist-3 quark-gluon operators has been studied in \cite{Beli1} with the emphasis on 
the solutions of evolution equations. 
In this letter we study the scale dependence of twist-3 quark-gluon matrix element $T_F(x_1,x_2)$ 
and $T_{\Delta,F}(x_1,x_2)$ with $x_{1,2} >0$ with a different method. From our results one can obtain 
the scale dependence in the special case of $x_1=x_2$ or of $x_{1,2}=0$.  
\par 
To study the scale dependence of the matrix elements defined with twist-3 quark-gluon operators, it is convenient 
to introduce as in \cite{BMP}:
\begin{equation}
  T_{\pm} (x_1,x_2) = T_F(x_1,x_2) \pm T_{\Delta,F} (x_1,x_2). 
\end{equation}
The advantage for working with $T_{\pm}$ is that the non-singlet parts of $T_{\pm}$ does not mix under 
renormalization. 
The singlet parts are certainly mixed with the purely gluonic matrix elements. We only need to consider the scale dependence of $T_+$. The scale dependence 
of $T_-$ can be obtained from the properties given 
in Eq.(\ref{TFSY}). 
At one-loop level, we can divide the scale-dependence or the evolution into three parts:
\begin{equation} 
\frac{\partial T_{+}(x_1,x_2,\mu)}{\partial \ln\mu} = 
\frac{\partial T_{+}(x_1,x_2,\mu)}{\partial \ln\mu} \biggr\vert_{qg}  
+\frac{\partial T_{+}(x_1,x_2,\mu)}{\partial \ln\mu} \biggr\vert_{q\bar q} 
+ \frac{\partial T_{+}(x_1,x_2,\mu)}{\partial \ln\mu} \biggr\vert_{gg}. 
\label{EVO3}  
\end{equation}  
In the above the first two parts are of the non-singlet part. The first part is a convolution 
of $T_+(x_1,x_2)$ with $x_{1,2}>0$. The second part is a convolution with $T_+(x_1,x_2)$ 
with $x_1 <0$ or $x_2 <0$. The third part is the mixing part with the purely gluonic matrix elements.   
Detailed expressions of each part will be given. 
\par 
Before we study the scale dependence from each part we briefly explain our method. 
Our method is to directly calculate these twist-3 matrix elements with parton states 
instead of a hadron state. 
If we use helicity instead of the spin vector $s$ to describe the spin, the above defined 
matrix elements with twist-3 operators, generically denoted as ${\mathcal O}$, 
are the non-diagonal parts of the matrix elements $\langle P,\lambda'\vert {\mathcal O }\vert P,\lambda\rangle$ 
in the $2\times 2$ helcity space. That is we need to study the forward scattering amplitudes
with helicity-flip. 
\par  
If we use a single quark to replace the hadron, we will always have $T_{\pm}=0$ 
because helicity conservation of QCD. However, we can use a multi-parton state 
instead of a single-quark state. Factorizations of SSA  
have been studied in \cite{MS1,MS2,MSZ,MZ} by using multi-parton states. 
We can introduce the following multi-parton state:
\begin{equation}
 \vert n [\lambda ] \rangle  =  \vert q(p,\lambda_q) [\lambda =\lambda_q  ] \rangle + {\mathcal C}^{qg} 
                   \vert q(p_1,\lambda_q) g(p_2,\lambda_g ) [\lambda =\lambda_q +\lambda_g ] \rangle +\cdots, 
\label{MPS}                    
\end{equation}
with $p_1+p_2 =p$. 
The $qg$-state is in the fundamental representation of $SU(N_c)$-gauge group. 
We take the momenta as $p_1=x_0 p$ and $p_2=(1-x_0) p$ 
with $p^\mu =P^\mu =(P^+, 0,0,0)$. ${\mathcal C}^{qg}$ is a real constant. 
If we calculate $T_{\pm}$ with this state in Eq.(\ref{tw3}), 
we will find that $T_{\pm} $ receives nonzero contributions only from the matrix elements 
of the interference between the single quark- and the $qg$-state, i.e., $\langle q(\lambda_q)\vert {\mathcal O }
\vert q(\lambda_q)g(\lambda_g)\rangle$ or $\langle q(\lambda_q)g(\lambda_g) \vert {\mathcal O }
\vert q(\lambda_q)\rangle$. It is noted that the total helicity in the bra- and ket-state is different, 
but the quark always has the same helicity. At tree-level, we obtain: 
\begin{eqnarray}
T_+^{(0)} (x_1,x_2) &=&  g_s  \pi {\mathcal C}^{qg}  \sqrt{2 x_0} (N_c^2-1)(x_2-x_1)
                      \delta (1-x_1) \delta (x_2-x_0 ),
\nonumber\\
T_{-}^{(0)}(x_1,x_2) &=&  - g_s \pi {\mathcal C}^{qg}  \sqrt{2x_0 } (N_c^2-1)(x_2-x_1)
  \delta (1-x_2) \delta (x_1-x_0).
\label{TreeQG}
\end{eqnarray}
Because the multi-parton state depends on $x_0$ with $x_0 <1$ per definition, $T_{\pm}$ 
calculated with the state will also depend on $x_0$. But the scale dependence or the evolution 
will not depnd on $x_0$ and ${\mathcal C}^{qg}$.   
\par 
It is possible to have more multi-parton states in Eq.(\ref{MPS}), represented with $\cdots$.
However, not all possible multi-parton states for the cases with twist-3 operators 
are needed\cite{MSZ}. 
The basic idea for cases with twist-3 operators is to consider  
various helicity-flip matrix elements like $\langle a,b\vert {\mathcal O}\vert c \rangle$ 
and $\langle c \vert {\mathcal O}\vert a,b \rangle$ with $a,b$ and $c$ as possible partons in QCD.
Because of helicity flip we only need to consider those matrix elements with three combinations of partons:
$\langle q,g \vert {\mathcal O}\vert q \rangle$, $\langle q,\bar q \vert {\mathcal O}\vert g \rangle$  
and $\langle g,g  \vert {\mathcal O}\vert g \rangle$. The complex conjugated matrix elements should also 
be included for consistency. Because of the three combinations we can divide 
the evolution in Eq.(\ref{EVO3}) into three parts at one-loop.  For each combination with a given operator ${\mathcal O}$ 
one can construct the corresponding $2\times 2$ spin density 
matrix in helcity space for a spin-1/2 system. The non-diagonal part is relevant 
for the transverse polarization. For ${\mathcal O}$ being those operators 
used to define twist-3 matrix elements, one can extract these matrix elements from the 
corresponding non-diagonal parts of spin density matrices. The spin-density matrices for each combination 
are given in \cite{MSZ}. We will also call the contribution extracted from $\langle q,g \vert {\mathcal O}\vert q \rangle$, $\langle q,\bar q \vert {\mathcal O}\vert g \rangle$  
and $\langle g,g  \vert {\mathcal O}\vert g \rangle$ as $qg$-, $q\bar q$- and $gg$ contribution, respectively.
For detailed description of those parton states and spin-density matrices we refer to \cite{MSZ}. 
We will also use the same notations for these multi-parton states as used in \cite{MSZ}.

\par
\begin{figure}[hbt]
\begin{center}
\includegraphics[width=13cm]{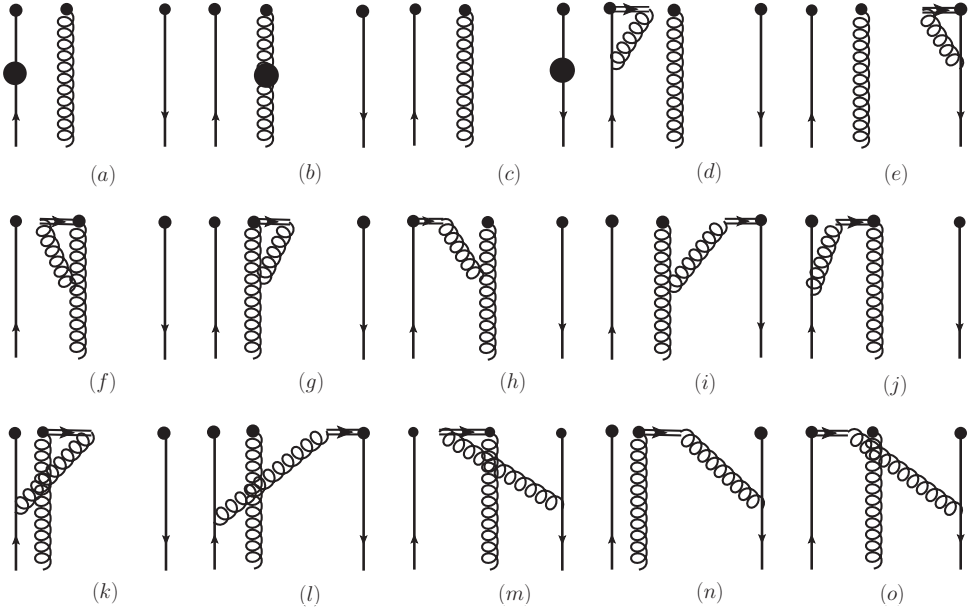}
\end{center}
\caption{A set of diagrams of one-loop corrections to $T_{\pm} (x_1,x_2)$ from the $qg$-contribution. This set
only contains the self-energy corrections represented by black dots,
 and corrections with one gluon emission from a gauge link. }
\label{P2}
\end{figure}
\par
\par 
Now we turn to the first part in Eq.(\ref{EVO3}). The scale dependence of this part 
can be written as the evolution: 
\begin{equation}
\frac{\partial \ T_{\pm} (x_1,x_2,\mu)}{\partial \ln\mu } \biggr\vert_{qg} = \frac{\alpha_s}{\pi}
\int_0^1  d\xi_1 d\xi_2 {\mathcal F}_{\pm}(x_1,x_2,\xi_1,\xi_2 ) T_{\pm} (\xi_1,\xi_2,\mu).
\label{RGT} 
\end{equation}
With the tree-level result given in Eq.(\ref{TreeQG}), the kernel ${\mathcal F}_+$ 
can be determined by calculating the one-loop correction of $T_+$.
The one-loop contributions, denoted as $T_+^{(1)}(x_1,x_2,x_0)$, are represented 
by diagrams given in Fig.1 and Fig.2. 
With the tree-level result and the scaling property of ${\mathcal F}_+$ we have 
at the leading order:
\begin{eqnarray} 
 g_s\alpha_s {\mathcal F}_+ (x_1, x_2,\xi_1,\xi_2) = \frac{1}{N_c^2-1}\frac{\xi_1}{\xi_2-\xi_1}  
   \sqrt{\frac{\xi_1}{2 \xi_2} } \frac{\partial}{\partial \ln \mu} T^{(1)}_+ (x_1/\xi_1,x_2/\xi_1, \xi_2/\xi_1) + \cdots,   
\label{derF}    
\end{eqnarray} 
where $\cdots$ denote the contribution from the $\mu$-dependence of $g_s$ appearing in the definition in Eq.(1). 
We notice here that the calculation of $T^{(1)}_+ (x_1/\xi_1,x_2/\xi_1, \xi_2/\xi_1)$ , hence of ${\mathcal F}_+$, is slightly different than that of $T^{(1)}_+(x_1,x_2,x_0)$. In the later, all variables $x_{0,1,2}$
are always smaller than 1. Especially, with $x_0 < 1$ one finds that $T_+(x_1,x_2,x_0)$ are nonzero 
only for $x_{1,2}<1$ and $\vert x_1-x_2\vert <1$. 
While in the former, any of the variables in $T^{(1)}_+ (x_1/\xi_1,x_2/\xi_1, \xi_2/\xi_1)$ can be larger than $1$.
We take Fig.1h as an example to explain the difference. 
\par 
We denote the momentum of the gluon emitted from the gauge link as $k$, the contribution to $T_+^{(1)}(x_1,x_2,x_0)$  
from Fig.1h can be written in the form:
\begin{equation} 
  T_+ (x_1,x_2,x_0)\biggr\vert_{1h} \propto \delta (1-x_1) \mu^{\epsilon} \int \frac{ dk^- d^{d-2} k_\perp }{(2\pi)^3} 
     \frac{1}{ k^2 +i\varepsilon} \frac{1}{ (p_2-k)^2 +i\varepsilon}  [ \cdots ]  
\end{equation}     
with $k^+$ fixed as $p_2^+ -k^+ =(1-x_2) p^+$. We use the dimensional regularization with $d=4-\epsilon$.
$\mu$ is the renormalization scale. 
There are two poles from the two propagators in the $k^-$-plane. 
Their positions are  
determined by $1-x_2$ and $x_2-x_0$, respectively. 
Performing the $k^-$- and $k_\perp$-integration one has the nonzero contribution proportional 
to  $\delta (1-x_1) \theta (x_2-x_0) \theta (1-x_2) $ to $T_+ (x_1,x_2,x_0)$. 
From the above expression for $T^{(1)}$ one can also obtain the kernel before the integrations by taking 
the derivative against $\ln \mu$. 
We have then
\begin{eqnarray} 
{\mathcal F}_+\biggr\vert_{1h} \propto && \lim_{\epsilon \to 0} \delta (1-x_1/\xi_1 ) \epsilon \mu^{\epsilon} \int \frac{ dk^- d^{d-2} k_\perp }{(2\pi)^3} 
     \frac{1}{ 2 (x_2/\xi_1 -\xi_2/\xi_1) p^+ k^-  -k_\perp^2  +i\varepsilon} 
\nonumber\\     
    && \cdot  \frac{1}{ -2 (1-x_2/\xi_1) p^+ k^- -k_\perp^2  +i\varepsilon}  [ \cdots ]. 
\end{eqnarray}   
Now we can see that the positions of the two poles are now determined by $(1-x_1/\xi_1)$ and 
$(x_2/\xi_1-\xi_2/\xi_1)$, respectively. Therefore, in the calculation of the kernel the positions 
of poles in the $k^-$-plane are differently determined than those in the calculation of $T_+ (x_1,x_2,x_0)$. 
It is then straightforward to obtain: 
\begin{eqnarray} 
{\mathcal F}_+ (x_1,x_2,\xi_1,\xi_2)\biggr\vert_{1h} &=& -\frac{N_c} {4}
\delta (\xi_1-x_1) \biggr ( \theta (x_2-\xi_2)  \theta(x_1-x_2)  - \theta (\xi_2-x_2) \theta (x_2-x_1) \biggr ) 
\nonumber\\
   && \cdot    \biggr [ \frac{x_1-x_2}{(x_1-\xi_2)^2} 
  -\frac{2}{x_2-\xi_2 } +\frac{2}{x_1-\xi_2} \biggr ] . 
\end{eqnarray}
We notice here that there is a singularity in the second term in $[ \cdots]$ for $\xi_2 \sim x_2$. 
This is a light-cone singularity which will be canceled. We also note that the first term is nonzero 
as a distribution for $x_1\to x_2$, although it is proportional to $x_1-x_2$.        
\par 
The result of Fig.1f reads:
\begin{equation} 
{\mathcal F}_+ (x_1,x_2,\xi_1,\xi_2)\biggr\vert_{1f} = -\frac{N_c}{2} \delta(\xi_1-x_1)\delta(\xi_2-x_2)
   \left  [   \int_0^{\vert x_1-x_2 \vert } \frac{dy}{y} -\frac{1}{2} \right ]. 
\end{equation} 
This contribution also contains a light-cone singularity. This will be canceled 
by that in the contribution from Fig.1h. In the final result there is no light-cone singularity. 
In calculating the above contribution a special care should be taken. At first look 
the integration variable $y$ may be changed by $y\to y/\vert x_1-x_2 \vert$ so that 
one obtains the integral over $y$ from $0$ to $1$ instead of $0$ to $\vert x_1-x_2 \vert$. 
But this change is not correct. The correct result is only obtained by keeping the interpretation 
of $y$ as a momentum fraction in unit of $p^+$. This can be shown as in the following: 
The origin of the integral 
comes from the integration over $k^+$ with the eikonal propagator $1/(k^+ +i\varepsilon)$, 
i.e., the propagator from the gauge link represented by the double line in Fig.1f.  
The light-cone singularity comes from the region with $k^+\sim 0$. 
One can regularized this singularity in such a propagator  
with the so-call $\eta$- or $\Delta$-regulator used in \cite{MII,CFHKM}.
With the regularization one can calculate all contributions involving the eikonal propagator. 
The results at end are the same as we have done for Fig.1f by keeping $y$ as the momentum 
fraction in unit of $p^+$.
\par 
We notice that the contribution from Fig.1b introduces an $N_F$-dependence in the kernel with $N_F$ 
as the number of quark flavors. By taking the scale dependence of $g_s$ in the definition in Eq.(\ref{tw3}) 
into account, the $N_F$-dependence is canceled. Adding all contributions 
from Fig.1 to ${\mathcal F}_+$ and the contribution from the scale dependence of $g_s$ in the definition  
we have: 
\begin{eqnarray} 
{\mathcal F}_+(x_1,x_2,\xi_1,\xi_2)\biggr\vert_{Fig.1}   &=& 
   \delta (\xi_1-x_1) \delta(\xi_2-x_2) \left ( -N_c \ln \vert x_1-x_2\vert +\frac{N_c^2-1}{N_c} \biggr ( \frac{3}{4} -\frac{1}{2} \ln x_2 
     -\frac{1}{2} \ln x_1 \biggr )  \right ) 
\nonumber\\   
     && + \frac{N_c}{4} \delta(\xi_1-x_1) \biggr ( \theta(x_1-x_2) \theta(x_2-\xi_2)
        -\theta(x_2-x_1) \theta (\xi_2-x_2) \biggr ) 
\nonumber\\        
      && \cdot    \left ( -\frac{x_1-x_2}{(x_1-\xi_2)^2} 
         +\frac{2}{(x_2-\xi_2)_+} -\frac{2}{x_1-\xi_2}  \right ) 
\nonumber\\
    && + \frac{N_c}{4} \delta (\xi_2-x_2) \biggr ( \theta (x_1-x_2)\theta (\xi_1-x_1) -\theta (x_2-x_1) \theta (x_1-\xi_1) \biggr ) 
\nonumber\\    
    && \cdot  \left ( -\frac{x_1-x_2}{(\xi_1-x_2)^2} 
       + \frac{2}{(\xi_1-x_1)_+} -\frac{2}{\xi_1-x_2}   \right ) 
\nonumber\\
    && + \frac{N_c}{2}\biggr [ \delta (\xi_1-x_1) \theta (\xi_2-x_2) \frac{x_2}{\xi_2} \biggr ( 
         \frac{1}{(\xi_2-x_2)_+ } -\frac{1}{\xi_2-x_1} \biggr ) 
\nonumber\\         
    &&    + \delta (\xi_2-x_2) \theta (\xi_1-x_1) \frac{x_1}{\xi_1} \biggr ( \frac{1}{(\xi_1-x_1)_+} -\frac{1}{\xi_1-x_2} \biggr ) 
        \biggr ]  
\nonumber\\
   && - \frac{1}{2 N_c} \delta (x_1-x_2-\xi_1+\xi_2) \theta (\xi_2-x_2) 
        \biggr ( \frac{x_2}{\xi_2} +\frac{x_1}{\xi_1} \biggr ) \frac{1}{(\xi_2-x_2)_+} .             
\end{eqnarray}
The $+$-distributions appearing in the above are defined as:
\begin{eqnarray} 
\int_0^1  d \xi  \frac{ \theta (x-\xi) f(\xi)}{(x-\xi)_+} &=& \int_0^x d \xi \frac{f(\xi)-f(x)}{x-\xi } + f(x) \ln x ,  
\nonumber\\
\int_0^{1} d \xi \frac{\theta (\xi-x) f(\xi)}{(\xi-x )_+} &=& \int_x^1 d \xi \frac{f(\xi)-f(x)}{\xi-x} + f(x) \ln (1-x) .
\end{eqnarray}
These $+$-distributions are different than the standard $+$-distribution which will be given later.         

\par 
\par\vskip20pt  
\par
\begin{figure}[hbt]
\begin{center}
\includegraphics[width=12cm]{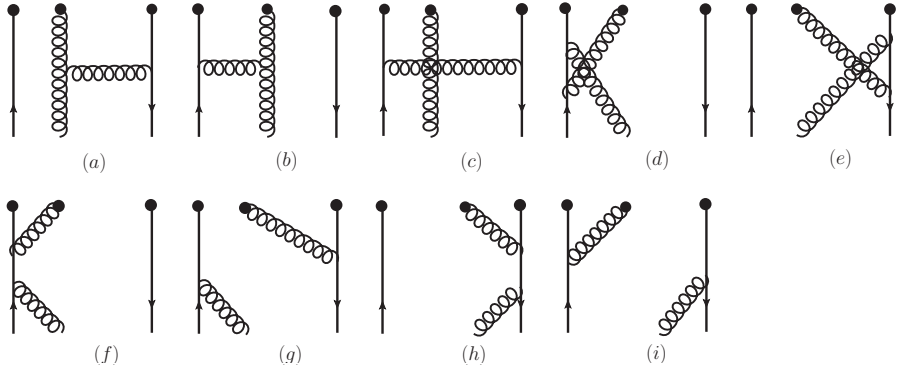}
\end{center}
\caption{Another set of diagrams for one-loop corrections of $T_{\pm}$ from the $qg$-contribution.   }
\label{SFP-P6}
\end{figure}
\par

\par
The contributions from Fig.2 can easily be worked out. The results are: 
\begin{eqnarray}
{\mathcal F}_+ (x_1,x_2,\xi_1,\xi_2) \biggr\vert_{2a} &=& \frac{N_c}{4} \frac{\delta(\xi_2-x_2)}{\xi_2-\xi_1} 
 \biggr [ \theta (\xi_1-x_1) \biggr ( 
     - 2 \frac{x_1}{\xi_1} \theta (x_2-x_1) 
\nonumber\\
    && + \frac{\theta(x_1-x_2)}{\xi_1-\xi_2} \biggr ( 2 \frac{x_1}{\xi_1} x_2-x_2-x_1   \biggr ) \biggr ) -\theta (x_1-\xi_1) \theta (x_2-x_1) \frac{x_1-x_2}{\xi_1-x_2} \biggr ] ,
\nonumber\\
{\mathcal F}_+ (x_1,x_2,\xi_1,\xi_2) \biggr\vert_{2b} &=& - \frac{N_c}{4}\delta (\xi_1-x_1)
  \biggr [ \frac{\theta (x_2-x_1) \theta (\xi_2-x_2)}{\xi_2 (\xi_2-x_1)^2} 
      ( 2x_2^2-2x_1x_2 -2\xi_2^2 -\xi_2x_2 +3x_1\xi_2 ) 
\nonumber\\
  && + \frac{\theta (x_1-x_2)}{\xi_2-x_1} \biggr ( \frac{2x_2}{x_1\xi_2} (x_1-x_2-\xi_2) \theta (\xi_2-x_2) 
       +\frac{x_1-x_2}{x_1-\xi_2} \theta (x_2-\xi_2) \biggr ( 3 -2\frac{x_2}{\xi_1} -2 \frac{\xi_2}{\xi_1}\biggr ) \biggr ],       
\nonumber\\
{\mathcal F}_+ (x_1,x_2,\xi_1,\xi_2) \biggr\vert_{2c} &=& -\frac{1}{2N_c} \delta (x_1-x_2-\xi_1+\xi_2)
\theta (\xi_2-x_2) \frac{\xi_1-x_1}{\xi_1 \xi_2},
\nonumber\\
{\mathcal F}_+ (x_1,x_2,\xi_1,\xi_2) \biggr\vert_{2d} &=& \frac{\delta(\xi_1-x_1)}{2 N_c} \frac{x_2-x_1 +\xi_2}{x_1 (\xi_2-x_1)} 
 \theta (x_1-x_2) 
\nonumber\\ 
   && \left ( \frac{x_1-x_2}{\xi_2}\theta (x_2-x_1+\xi_2) +\frac{x_2}{x_1-\xi_2}\theta(x_1-x_2-\xi_2) \right ),    
\nonumber\\
{\mathcal F}_+ (x_1,x_2,\xi_1,\xi_2) \biggr\vert_{2f}  &=& \frac{N_c^2-1}{2N_c} \delta (\xi_1-x_1) \theta (x_1-x_2) 
\frac{x_2(x_1-x_2)}{x_1^2 (\xi_2-x_1)},            
\label{F3abc} 
\end{eqnarray}
the contributions from Fig.2e, 2g, 2h and 2i are zero. We do not try to give the sum these contributions because 
the result is too lengthy. Our final result for ${\mathcal F}_+$, i.e, the first part is then given by Eq.(\ref{RGT}) 
with the kernel:
\begin{equation} 
  {\mathcal F}_+ = {\mathcal F}_+\biggr\vert_{Fig.1} +\sum_{i=a,b,c,d,f} {\mathcal F}_+\biggr\vert_{2i}. 
\end{equation} 
In this part it is trivial that the kernel does not depend on ${\mathcal C}^{qg}$.  
\par
\begin{figure}[hbt]
\begin{center}
\includegraphics[width=9cm]{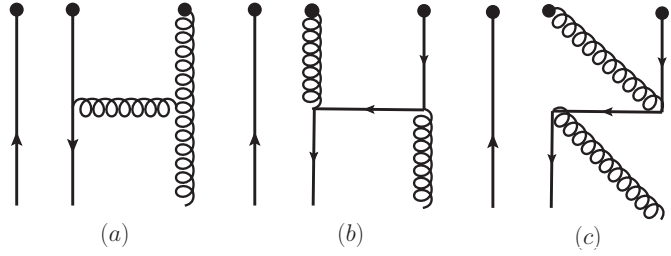}
\end{center}
\caption{One-loop diagrams in the light-cone gauge for $T_{\pm}(x_1,x_2)$ from the $q\bar q$-contribution. }
\label{qqbar}
\end{figure}       
\par 
Now we turn to the second part or the $q\bar q$-contribution in Eq.(\ref{EVO3}). As discussed before, in this case we need to consider the matrix element 
as $\langle q (p_1) ,\bar q (p_2) \vert {\mathcal O}\vert g (p) \rangle$, where momenta 
of partons are specified.  In constructing the corresponding spin density matrices
for this part, one notes that the $q\bar q$-state must have the total helicity $\lambda =0$. 
There are two possible $q\bar q$-states with $\lambda=0$. One is the symmetric state in helicity state. Another one 
is antisymmetric state. Therefore, the general $q\bar q$ -state is a superposition 
of these two states. We denote the weight of the symmetric state in the superposition as ${\mathcal C}_+^{q\bar q}$, and the 
weight of the anitysmmetric state as ${\mathcal C}_-^{q\bar q}$. The $q\bar q$ is in color-octet. 
The tree-level result for the $q\bar q$-contribution reads:
\begin{eqnarray}
T_+^{(0)} (x_1,x_2) =  \pi g_s  (N_c^2-1) \sqrt{2x_0\bar x_0} \biggr [ 
 ({\mathcal C}_+^{q\bar q}-{\mathcal C}_-^{q\bar q}) \delta (x_1+\bar x_0) \delta (x_2-x_0)
   + ({\mathcal C}_+^{q\bar q}+ {\mathcal C}_-^{q\bar q})   \delta (x_2 +\bar x_0) \delta (x_1-x_0) \biggr ].
\end{eqnarray}
At tree-level, $T_+(x_1,x_2)$ of the $q\bar q$-contribution is zero for $x_{1,2}>0$, but 
nonzero for $x_1 <0$ or $x_2 <0$. 

\par 
At one-loop, $T_+(x_1,x_2)$ with $x_{1,2}>0$ can be nonzero. It receives contributions from diagrams in Fig.3. 
In Fig.3 the diagrams are for contributions in the light-cone gauge $n\cdot G=0$. In Feynman gauge there are more 
diagrams. The scale dependence of this part can be written as the convolution:
\begin{eqnarray}
\frac{\partial \ T_{+} (x_1,x_2,\mu)}{\partial \ln\mu }\biggr\vert_{q\bar q}  &=& \ \frac{\alpha_s}{\pi}
\int_0^1   d\xi_1 d\xi_2 {\mathcal F}_{+1}(x_1,x_2,\xi_1,\xi_2 ) \theta (\xi_1-\xi_2) T_{+} (\xi_2-\xi_1,\xi_2) 
\nonumber\\
  && +\frac{\alpha_s}{\pi}
\int_0^1  d\xi_1 d\xi_2 {\mathcal F}_{+2}(x_1,x_2,\xi_1,\xi_2 ) \theta (\xi_2-\xi_1)  T_{+} (\xi_1,\xi_1-\xi_2).  
\end{eqnarray}
In the above we have $\xi_{1,2}>0$. From Fig.3 we can obtain the kernels as:
\begin{eqnarray} 
{\mathcal F}_{+1} (x_1,x_2,\xi_1,\xi_2)  &=& \frac{N_c}{2}\delta (\xi_2-x_2) \theta(x_2-x_1)  
       \frac{(x_1-x_2)^2}{\xi_1^2 (\xi_1-x_2+x_1)}
   -\frac{1}{2 N_c} \delta (\xi_2-x_2) 
  \theta (\xi_1-x_1)\frac{x_1}{\xi_1^2}, 
\nonumber\\
{\mathcal F}_{+2} (x_1,x_2,\xi_1,\xi_2) &=& -\frac{N_c}{2} \delta(\xi_1-x_1) \theta (x_1-x_2) 
\frac{ (x_2-x_1) (\xi_2^2 -x_2^2 + x_1x_2)}{\xi_2^2 \xi_1 ( \xi_2-x_1+x_2)}, 
\nonumber\\
  && +  \frac{1}{2N_c} \delta (\xi_1-x_1) \theta (\xi_2-x_2) 
    \frac{1}{\xi_2^2} 
  \biggr [ \theta (x_2-x_1) \frac{(\xi_2-x_2)^2}{\xi_2-x_1} + \theta (x_1-x_2) \frac{x_2 (\xi_2-x_2)}{x_1} \biggr]
\nonumber\\
   &&  + \frac{N_c^2-1}{2N_c}\delta(\xi_1-x_1) \theta(x_1-x_2) 
     \frac{x_2 (x_2-x_1)}{\xi_2 x_1^2}.
\label{QQBAR}          
\end{eqnarray} 
The derived kernels do not depend on the introduced weights ${\mathcal C}^{q\bar q}_\pm$ as expected. 
\par 
For the mixing part with the purely gluonic operators one needs to calculate the contribution $T_+$ from 
the matrix element like $\langle g(p_1),g(p_2)\vert {\mathcal O}\vert g(p) \rangle$. 
The state with the two gluons must have the same color of the state with one gluon. Therefore, 
the colors of the two gluons can be coupled through $if^{abc}$ and $d^{abc}$. 
The state with the two gluons 
must have the total helicity $\lambda =0$ for helicity flip. Again, there are two possible states. 
The general two gluon state with $\lambda =0$ is a superposition of the states. One is the symmetric state in helicity 
space, for which we introduce a weight ${\mathcal F}_+^{gg} $ for the color structure with $if^{abc}$. 
Another is the antisymmetric state in helicity space, for which we introduce 
a weight ${\mathcal F}_-^{gg}$ for the color structure with $if^{abc}$. One can introduce weights in a similar way 
for the color structure with $d^{abc}$. The corresponding spin density matrices for our calculation 
are defined in \cite{MSZ}. 
\par
\begin{figure}[hbt]
\begin{center}
\includegraphics[width=6cm]{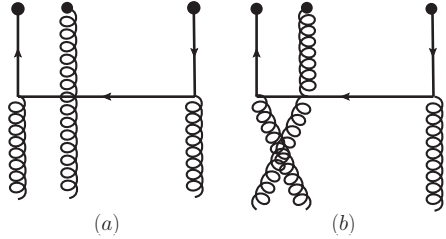}
\end{center}
\caption{The one-loop diagrams for  $T_{\pm}(x_1,x_2)$ from the $gg$-contribution. The gluons from the bottom 
are labeled as, the left gluon is with $p_1$, the middle one is with $p_2$, both are incoming. The right one 
is with $p$ as an outgoing gluon.  }
\label{Omix}
\end{figure}
\par 
\par 
At tree-level, we obtain the results for $T_G^{(f)}$ and $T_{\Delta,G}^{(f)}$ with $x_{1,2}>0$:
\begin{eqnarray} 
T^{(f,0)}_G(x_1,x_2) &=& \frac{ g_s \pi}{\sqrt{2}} N_c (N_c^2-1) x_1 x_2 (x_1-x_2) \biggr [ 
           C_+ \biggr (\delta (1-x_1) \delta(x_2-x_0) -\delta (1-x_2) \delta(x_1-x_0) \biggr ) 
\nonumber\\           
    && \ \ \ 
            - C_- \biggr ( \delta (1-x_1) \delta(x_2-\bar x_0) 
             -\delta (1-x_2) \delta(x_1-\bar x_0) \biggr ) \biggr ] ,            
\nonumber\\
T_{\Delta,G}^{(f,0)} &=& \frac{ g_s\pi}{\sqrt{2}} N_c (N_c^2-1) x_1 x_2 (x_1-x_2) \biggr [ 
           C_+ \biggr (\delta (1-x_1) \delta(x_2-x_0) +\delta (1-x_2) \delta(x_1-x_0) \biggr ) 
\nonumber\\           
    && \ \ \ 
            - C_- \biggr ( \delta (1-x_1) \delta(x_2-\bar x_0) 
             +\delta (1-x_2) \delta(x_1-\bar x_0) \biggr ) \biggr ] ,            
\nonumber\\
  C_+ &=& {\mathcal F}_+^{gg} + {\mathcal F}_-^{gg}, \ \ \ 
  C_-  =  {\mathcal F}_+^{gg} - {\mathcal F}_-^{gg}.
\end{eqnarray} 
For other cases, i.e., $x_{1,2} <0$ the functions can also be obtained.
Similar results for the color structure with $d^{abc}$ 
can be obtained with the replacement of color factors. 
At tree-level, $T_+$ is zero here, but nonzero at one-loop. At one-loop 
there are six diagrams which give contributions to $T_+$. Two of them are given
in Fig.4. Other four diagrams can be obtained through permutations.
Only the diagrams given in Fig.4 will contribute to the mixing part in the scale dependence for $x_{1,2}>0$.
The calculations are straightforward. We find that the contributions from the two diagrams 
are renormailized with the purely gluonic matrix elements with $x_{1,2}>0$. 
To give our result we introduce the notation:
\begin{equation} 
  T_{\pm G}(x_1,x_2) = \left ( T_G^{(f)}(x_1,x_2) + T_G^{(d)}(x_1,x_2) \right ) 
  \pm  \left (  T_{\Delta,G} ^{(f)}(x_1,x_2) 
   + T_{\Delta,G}^{(d)}(x_1,x_2) \right ). 
\end{equation} 
The mixing- or third part of the scale dependence in Eq.(\ref{EVO3}) can be given as:
\begin{eqnarray} 
\frac{\partial T_+(x_1,x_2,\mu) }{\partial \ln \mu } \biggr\vert_{gg} = \frac{\alpha_s}{\pi} 
\int_0^1  d\xi_1 d\xi_2   {\mathcal F}_{G}(x_1,x_2,\xi_1,\xi_2 ) T_{+G}(\xi_1,\xi_2),
\end{eqnarray}
with 
\begin{eqnarray}
   {\mathcal F}_{G} (x_1,x_2,\xi_1,\xi_2) &=& -\delta (x_1-x_2 -\xi_1 +\xi_2) \theta (\xi_1-x_1) 
       \frac{ x_1 x_2  }{2 \xi^2_2 \xi_1^2} 
\nonumber\\       
     &&  - \delta (x_2-x_1-\xi_1+\xi_2) \theta (\xi_1-x_2) 
     \frac{ (\xi_1-x_2)^2 }{2\xi_2^2\xi^2_1}.
\label{GG}     
\end{eqnarray}
Again, the kernel derived here does not depend on the states, i.e., on the weights like ${\mathcal F}^{gg}_{\pm}$.  
\par 
With the three parts in Eq.(\ref{EVO3}) given in the above, we have derived the scale dependence at one-loop for 
the twist-3 matrix element $T_{+}(x_1,x_2,\mu)$ for $x_{1,2}>0$. 
From the derived scale dependence one can easily find the scale dependence 
of $T_-(x_1,x_2,\mu)$ by using $T_-(x_1,x_2,\mu)=T_+(x_2,x_1,\mu)$. Two special cases can be derived from 
our general result. One is the evolution of the so-called soft-gluon-pole matrix element, which is obtained 
by taking $x_1 =x_2$. We can derive from Eq.(\ref{EVO3}) the evolution of $T_+(x,x,\mu)$ 
by taking the limits $x_1 \to x_2 \pm 0^+$. The results are same in these limits. 
We have with $z=x/\xi$: 
\begin{eqnarray}
\frac{\partial  T_F(x,x,\mu)}{\partial \ln \mu} &=& \frac{\alpha_s}{\pi} \biggr \{
    \int_x^1  \frac{dz}{z}   \biggr  [ P_{qq} (z) T_F(\xi,\xi)  + \frac{N_c}{2} 
    \frac{ (1+z) T_F(x,\xi) -(1+z^2) T_F(\xi,\xi) }{1-z}
     + T_{\Delta,F}(x,\xi) 
\nonumber\\
   && \ \  
+ \frac{1}{2N_c} \biggr (  (1-2z) T_{F}(x,x-\xi)
      + T_{\Delta,F} (x,x-\xi ) \biggr )  \biggr ]  -N_c T_F(x,x) 
\nonumber\\      
  &&  -\frac{1}{2}  \int_x^1  \frac{d z}{z}  \frac{ (1-z)^2 + z^2 }{\xi}    
    \left ( T_G^{(f)}(\xi,\xi) + T_G^{(d)}(\xi,\xi) \right )  \biggr \}, 
\label{SGP}     
\end{eqnarray}
where the quark splitting kernel is 
\begin{equation} 
  P_{qq} (z) = \frac{N_c^2-1}{2N_c}  \left [ \frac{1+z^2}{(1-z)_+} + \frac{3}{2} \delta (1-z) \right ]
\end{equation} 
and the standard $+$-distribution is defined as:
\begin{equation} 
   \int_x^1 d z \frac{f(z)}{(1-z)_+} = \int_x^1 \frac{f(z)-f(1)}{1-z} + f(1) \ln (1-x). 
\end{equation}
Since we have derived the same result in Eq.(\ref{SGP}) from the two limits $x_1 \to x_2 \pm 0^+$, it indicates that there is no scale dependence 
of $T_{\Delta,F}(x,x)$.    
Comparing existing results in \cite{KQ,BMP}, we find that our result of the non-singlet part, i.e., the contributions 
in the first- and second line in Eq.(\ref{SGP}), is in agreement with that in \cite{BMP}. The main difference between the results 
in \cite{KQ,BMP,VoYu,ZYL} is the last term in the second line. We notice that during preparing our work the last term is also 
confirmed in \cite{ZhSc}. The mixing part is in agreement with that given in \cite{KQ} by taking the difference 
of definitions into account. Our method for deriving the scale dependence is different 
than that in \cite{KQ,BMP,VoYu,ZYL,ZhSc}. Therefore, our result for the soft-gluon-pole matrix element 
gives an independent verification of existing results.  
\par
Another special case is the evolution of soft-quark-pole matrix elements. In this case one has either $x_1=0$ or $x_2=0$. 
Form our general results we can derive:
\begin{eqnarray} 
\frac{\partial  T_+(0,x,\mu)}{\partial \ln \mu}   &=& \frac{\alpha_s}{\pi} \biggr \{ \int_{x}^1 \frac{d z}{z}  \biggr [ -\frac{1}{2N_c} 
\frac{T_+(\xi-x,\xi)}{(1-z)_+} +\frac{N_c}{2} \frac{1+ z^3}{(1-z)_+}  T_+ (0,\xi) \biggr] 
\nonumber\\
   && + \frac{3 (N_c^2 -1)}{4N_c}   T_+(0,x)
 +  \int_x^1 \frac{dz}{z}  \biggr [ \frac{N_c}{2} \frac{z^2}{ (1-z)_+ } T_+(x-\xi,x) +\frac{1}{2N_c} (1-z)^2 T_+(0,-\xi)\biggr ]       
\nonumber\\
   &&  -\frac{1}{2x} T_{+G} (0,x) -\frac{1}{2}\int_x^1 \frac {d z}{z \xi} T_{+G}(\xi,\xi-x) \biggr \},      
\nonumber\\
\frac{\partial  T_+(x,0,\mu)}{\partial \ln \mu}    &=& \frac{\alpha_s}{\pi} \biggr \{ \int_{x}^1 \frac{dz}{z}  \biggr [
  -\frac{1}{2N_c} \frac{T_+(\xi,\xi-x)}{(1-z)_+} + \frac{1}{2N_c} T_+(x,\xi) +\frac{N_c}{2}\frac{z(1+z)}{(1-z)_+} 
    T_+ (\xi,0) \biggr ]
\nonumber\\
   && + \frac{3 (N_c^2-1) }{4 N_c} T_+ (x,0) 
 + \int_x^1 \frac{dz}{z}  \biggr [ \frac{N_c}{2} \frac{1}{(1-z)_+ } T_+(x,x-\xi) -\frac{1}{2N_c} z T_+(-\xi,0) \biggr ]     
\nonumber\\
  &&  -\frac{1}{2x} T_{+G} (x,0) -\frac{1}{2}\int_x^1 \frac {d z}{z \xi} T_{+G}(\xi-x, \xi) \biggr \},       
\label{SFP}
\end{eqnarray}
It is interesting to note that the first- and second part in Eq.(\ref{EVO3}) 
are not finite for $x_1=0$ or $x_2=0$. They contain light-cone singularities. But they are canceled in the sum.                      
\par 
To summarize: We have derived the evolution of the matrix elements of twist-3 quark-gluon operators. These matrix elements 
are important ingredients for predict SSA in QCD collinear factorization. From our results 
we have derived evolutions in two special cases. One is of soft-gluon-pole matrix elements, another is of soft-quark-pole 
matrix elements.  A comparison has been made with the existing results for soft-gluon-pole matrix elements. 
In this letter, we have only considered the case with $x_{1,2}>0$ in the evolution. The study of other cases 
including the evolution of purely gluonic matrix elements are in progress.

\par\vskip20pt
\noindent
{\bf Acknowledgments}
\par
We thank Dr. J.W. Qiu and Dr. F. Yuan for interesting discussions. We also thank Dr. H.Z. Sang and Mr. S.J. Zhu 
for the collaboration at the early stage of the work. The work of J.P. Ma is supported by National Nature
Science Foundation of P.R. China(No. 10975169,11021092), and the work of Q. Wang is supported by National Nature
Science Foundation of P.R. China(No.10805028),

\par

\end{document}